\documentclass[fleqn,usenatbib]{mnras}

\pdfoutput=1

\usepackage{newtxtext,newtxmath}
\usepackage[T1]{fontenc}

\DeclareRobustCommand{\VAN}[3]{#2}
\let\VANthebibliography\thebibliography
\def\thebibliography{\DeclareRobustCommand{\VAN}[3]{##3}\VANthebibliography}

\usepackage{graphicx}	% Including figure files
\usepackage{amsmath}	% Advanced maths commands
\usepackage {threeparttable}

\def\endash{\text{--}}

\newcommand{\DM}{\mathrm{DM}}
\newcommand{\GC}{\mathrm{GC}}

\newcommand{\f}[2]{\frac{#1}{#2}}
\newcommand{\powten}[1]{\times 10^{#1}}
\newcommand{\Rgc}{R_\mathrm{GC}}

\newcommand{\kpc}{\,\mathrm{kpc}}
\newcommand{\Mpc}{\,\mathrm{Mpc}}
\newcommand{\msun}{\,\mathrm{M}_\odot}
\newcommand{\Myr}{\,\mathrm{Myr}}
\newcommand{\Gyr}{\,\mathrm{Gyr}}

\def\runzero{\textit{Run0}}
\def\runone{\textit{Run1}}

\defcitealias{Ogiya_main}{O22}

\title[Impact of Dynamical Friction on the Tidal Formation of NGC 1052--DF2]{Impact of Dynamical Friction on the Tidal Formation of NGC 1052--DF2}

\author[Katayama, Nagamine, \& Kihara]{
Ryosuke Katayama$^{1}$,
Kentaro Nagamine$^{1,2,3,4,5}$\thanks{E-mail: kn@astro-osaka.jp (KN)},
Kenji Kihara$^{1}$
\\
$^{1}$ Theoretical Astrophysics, Department of Earth and Space Science, Graduate School of Science, Osaka University, Toyonaka, Osaka 560-0043, Japan\\
$^{2}$ Theoretical Joint Research, Forefront Research Center, Graduate School of Science, Osaka University, Toyonaka, Osaka 560-0043, Japan\\
$^{3}$ Kavli IPMU (WPI), The University of Tokyo, 5-1-5 Kashiwanoha, Kashiwa, Chiba, 277-8583, Japan \\
$^{4}$ Department of Physics \& Astronomy, University of Nevada, Las Vegas, 4505 S. Maryland Pkwy, Las Vegas, NV 89154-4002, USA\\
$^{5}$ Nevada Center for Astrophysics, University of Nevada, Las Vegas, 4505 S. Maryland Pkwy, Las Vegas, NV 89154-4002, USA \\
}

\date{Accepted XXX. Received YYY; in original form ZZZ}

\pubyear{2023}

\begin{document}
\label{firstpage}
\pagerange{\pageref{firstpage}--\pageref{lastpage}}
\maketitle

% Abstract of the paper (not more than 250 words for Main Journal papers or 200 words for Letters)
\begin{abstract}
The formation of dark matter-deficient galaxies (DMDGs) through tidal interactions has been a subject of growing interest, particularly with the discovery of galaxies such as NGC 1052--DF2. Previous studies suggested that strong tidal forces could strip dark matter from satellite galaxies, but the role of dynamical friction in this process has been largely overlooked.
In this paper, we present self-consistent N-body simulations that incorporate the effects of dynamical friction on the tidal formation of DF2, and compare them with the one without dynamical friction.
We find that dynamical friction significantly accelerates the decay of the satellite galaxy’s orbit, causing it to experience more frequent tidal stripping and leading to the earlier formation of a DM-deficient state, approximately $7-8$\,Gyr after infall. 
This is a few Gyr earlier than simulations without dynamical friction. 
Our results suggest that DMDGs can form in a wider range of orbital configurations, particularly on more circular orbits, than previously thought. 
Furthermore, we find that globular clusters in the DM-deficient phase exhibit elevated velocity dispersion, providing an observational signature of this evolutionary stage. 
We also examine the evolution of satellite in the phase space of total energy versus angular momentum, and show that a vertically narrow feature in this phase space is a clear signature of pericentre passage. 
These findings broaden the understanding of how DMDGs form and highlight the critical role of dynamical friction in shaping the evolutionary history of satellite galaxies in massive halos.
\end{abstract}

% \begin{keywords} % These are AAS keywords
% Dark matter(353) -- Dynamical friction(422) -- Galaxy formation(595) -- Galaxy interactions(600) -- Low surface brightness galaxies(940) -- Tidal interaction(1699)
% \end{keywords}

%%%%%%%%%%%%%%%%%%%%%%%%%%%%%%%%%%%%%%%%%%%%%%%%%%

%%%%%%%%%%%%%%%%% BODY OF PAPER %%%%%%%%%%%%%%%%%%

\section{Introduction} \label{sec:intro}

Since the discovery of NGC 1052--DF2 (hereafter DF2) as a dark matter-deficient galaxy (DMDG) by \citet{DF2_Nature}, the formation mechanisms of such galaxies have been widely debated. 
Shortly after its discovery, the accuracy of the distance measurement to DF2 was questioned by \citet{Trujillo2019}.  However, more recent data, based on 40 orbits of Hubble Space Telescope (\textit{HST}) Advanced Camera for Surveys (\textit{ACS}) observations, supports a distance of approximately $22\Mpc$  \citep{Shen2021b}.
The stellar mass of DF2 is $M_* \approx 2\powten{8}\msun$ 
\citep{DF2_Nature}, and 
its dynamical mass is estimated at $M_\mathrm{dyn} \approx 3\powten{8}\msun$, though with considerable uncertainties ($\approx 40\endash70$\%) \citep{DF2_Nature, Danieli2019, Emsellem2019}.

The distribution of globular clusters (GCs) around DF2 has been controversial.
There are about ten GCs around DF2 \citep{DF2_Nature, Shen2021a} with peak luminosities around $M_{606}\approx -9~\mathrm{mag}$ in F606W of the \textit{ACS} \citep{Shen2021a}, which is brighter than those in normal dwarf galaxies ($M\approx -7.5~\mathrm{mag}$) with similar stellar masses 
\citep{Miller&Lotz2007, Jordan2007, Rejkuba2012, Saifollahi2022}.
Additionally, the GCs exhibit a broad spatial distribution, spanning a radial distance of $1-10$\,kpc from DF2’s center \citep{DF2_Nature, Shen2021a}, which contrasts with the typically more centrally concentrated GC systems in early-type dwarf galaxies \citep{Tremaine1975, Antonini2013, NSCreview_Neumayer2020}.

There is another DMDG in the neighbourhood: NGC 1052--DF4 (hereafter DF4).
Surprisingly, DF2 and DF4 are quite similar to each other in that their age, mass, size, number of GCs, the spatial distribution of GCs, and the peak of the luminosity function of GCs
\citep{DF4_discovery, Danieli2020, Shen2021a}.
This similarity raises the question of whether DF2 and DF4 share a common formation pathway or if their resemblance is coincidental.
The possibility of forming DMDGs like DF2 and DF4 within the $\Lambda$ cold dark matter ($\Lambda$CDM) paradigm has been debated, with some concerns over whether current velocity measurements of the GC dynamics are accurate enough to support a truly DM-deficient nature \citep{Martin2018,Lewis2020} and the lack of observed tidal tails \citep{Montes2021}.

Among the proposed formation scenarios for DF2, the ‘mini-bullet cluster’ scenario has gained attention \citep{Silk2019, Shin2020, Lee2021, Otaki2022}.
This scenario suggests that DF2 formed from the collision of two gas-rich dwarf galaxies. 
While \citet{vanDokkum2022a, vanDokkum2022b} found that some characteristics of DF2 and DF4 align with this scenario, it struggles to explain the wide spatial distribution of GCs in DF2. 
Dark matter and stellar densities of DMDGs are sparse, so dynamical friction will be weak, and core stalling \citep{Read2006} may prevent them from sinking \citep{Dutta_Chowdhury2019, Dutta_Chowdhury2020}.
Even if this is the case for DF2, it does not mean that the GCs will be pushed to the outer region.
In fact, \citet{Ogiya2022} showed that the initial distribution of the GCs at their formation epoch is $r \approx 5\endash10\kpc$, and that eventually it is difficult for the mini-bullet cluster scenario to reproduce the spatial distribution of GCs of DF2.

An alternative explanation is the tidal formation scenario  \citep{Ogiya2018, Jackson2021, Montes2020, Moreno2022,Zemaitis23}, where a satellite galaxy orbiting a massive host halo experiences strong tidal forces. 
The tidal force inflates the satellite system (tidal heating) and even removes the mass in its outer region (tidal stripping).
The stars of the satellite galaxy are more concentrated in the center than the DM, so the tidal interaction causes a satellite to lose more DM than stars until when the system is stripped to an extent that the tidal radius is comparable to the stellar extent.
\citet[hereafter \citetalias{Ogiya_main}]{Ogiya_main} showed that the wide GC distribution in DF2 could  be explained by the tidal scenario using N-body simulations.
However, \citetalias{Ogiya_main} ignored the impact of dynamical friction on the satellite galaxy orbiting in the density field of the host halo by employing a time-varying analytical potential of the host halo. 

In addition to DF2 and DF4, we also make note of another ultra-diffuse galaxy F8D1 \citep{Caldwell98} associated with M81, the closest known example of such a kind. 
\citet{Zemaitis23} discovered a giant tidal tail of stars associated with F8D1, with an average surface brightness of $\mu_g \sim 32$\,mag\,arcsec$^{-2}$, extending over 60\,kpc in a projected distance.  They estimated that $30-36$ per cent of its current luminosity is contained in the tail, and that the close encounter with M81 is the likely cause of its dynamical feature.  Their discovery of the F8D1 tidal tail suggests that many other ultra-diffuse galaxies could be the result of such tidal interaction, and their tidal tail signatures could be hidden below the current detection limits. 
In Appendix~\ref{app:SB}, we show the surface brightness distribution of our simulated satellite galaxy. 

In this paper, we investigate the effect of dynamical friction on the tidal formation scenario of DMDGs, based on completely self-consistent N-body simulations.
Throughout this paper, we assume the distance to DF2 is $D = 20\Mpc$.
%
%% Summary of later parts
The rest of this paper is organised as follows. 
Section~\ref{sec:simulation} describes our simulations.
Section~\ref{sec:results} explains our analysis method and results.
Section~\ref{sec:dynamicalFriction} discusses the major channel for dynamical friction to make the resultant differences, and Section~\ref{sec:comp_ogiya} compares our results to those in the literature.
Finally, we summarise in Section~\ref{sec:conclusion}.

\section{Numerical Simulations and Setup} \label{sec:simulation}

We simulate a system consisting of a host halo and a satellite galaxy.
The tidal interaction causes the satellite to diffuse and lose its mass, eventually making it DM-deficient.

We run two types of simulations: 
`\runzero' has an N-body satellite galaxy and an external static potential as a host halo, 
and `\runone' has an N-body satellite and an N-body host halo.
This is the only difference between the two.
In {\runzero}, the satellite galaxy does not feel dynamical friction, while one does in \runone.
This way, we can clearly demonstrate the impact of dynamical friction by comparing the two runs.

We use the N-body/smoothed particle hydrodynamics (SPH) code \textsc{gadget3-osaka} \citep{gadget3osaka,Nagamine21}, which is a modified version of \textsc{gadget-2} \citep{Gadget2_Springel2005}. 
The code includes various physical modules such as star formation, supernova feedback, radiative heating and cooling, etc., which are all turned off in the present work because we are not doing galaxy formation simulations here. 
 The gravity calculation scheme is the tree method 
\citep{TreeMethod}, and we turn off the SPH part for this work. 
For the satellite, we placed 25 million DM particles and 49,990 stellar particles, whose masses are $4.00\times 10^3 \msun$. 
For the host halo in \runone, we put 15 million DM particles whose mass is $5.78\powten{5}\msun$. 
The Plummer softening length is $14~\mathrm{pc}$
\citep{PlummerSoftening}. 
The tree opening angle criterion parameter is set to $\theta = 0.6$.

\subsection{Initial condition} \label{sec:IC}

\begin{figure}
  \centering
    \includegraphics[keepaspectratio,width=0.9\columnwidth]{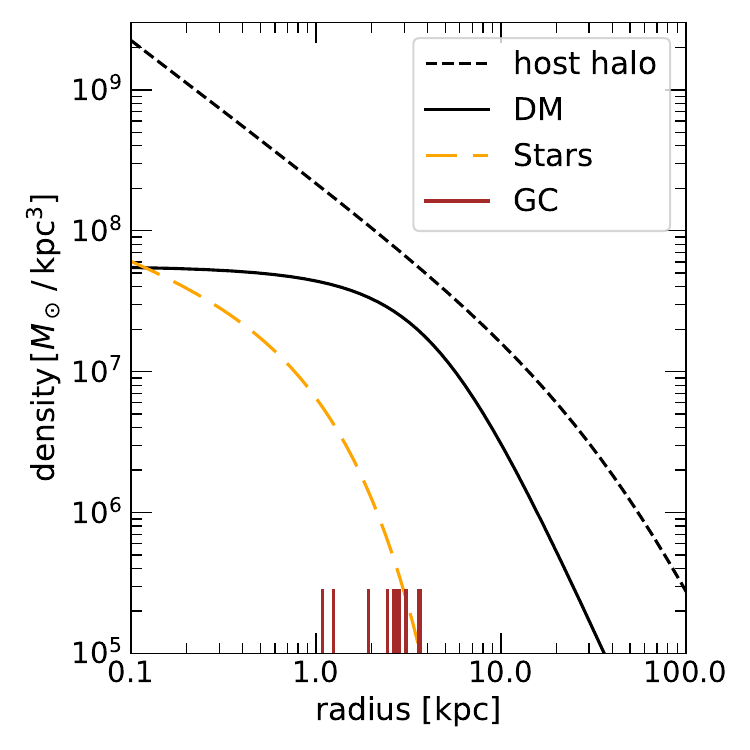}
  \caption{Initial density profiles of each component are shown for satellite DM (black solid), stars (orange long dashed), and host halo (black dashed). 
    Note that in {\runzero}, the host halo is implemented as an external potential, so there is no density profile of the host halo represented by particles. The bottom vertical lines indicate the initial radial positions of the ten globular clusters.
  }
  \label{fig:initialDensityProfile}
\end{figure}

We create the initial conditions similar to those in \citetalias{Ogiya_main}, 
which allows us to compare our results with theirs. 
% Host components
The host halo consists of only dark matter.
Its density profile is described by NFW profile \citep{NFW}:
\begin{equation}
  \rho(r) = \frac{\rho_s}{\frac{r}{r_s} \left(1+\frac{r}{r_s}\right)^2}, 
\end{equation}
where $\rho_s$ and $r_s$ are the scale density and scale radius, respectively.
The age of DF2 is estimated to be $\sim 9\Gyr$ \citep{vanDokkum2018b, Fensch2019, Ruiz-Lara2019}, 
which corresponds to redshift $z \sim 1.5$.
Therefore, we used $(M_{200},\, c) = (7\powten{12}\msun,\, 4.2)$ for the host halo based on \citet{Ludlow2016} assuming $z=1.5$.
{\runzero} implements the host halo as an external static potential, while {\runone} has an N-body one, although both of them use the same settings as written above.
In other words, {\runone} just calculates the gravity among host and satellite particles, while {\runzero} first calculates the gravity among satellite particles and then adds the analytical forces by the potential to them.

% Satellite stars
The satellite galaxy consists of dark matter, stars, and GCs.
The total stellar mass is set to $M_* = 2\powten{8}\msun$.
The stellar particles are distributed as the 3D-deprojected S\'{e}rsic profile 
\citep{Sersic, Sersic2, Sersic3, Sersic4}:
\begin{equation}
    \rho(r) = 
    \rho_s \left(\f{r}{R_e}\right)^{-p_n} \exp\left[-b_n \left(\f{r}{R_e}\right)^{1/n}\right]\text{,}
\end{equation}
\begin{equation}
    p_n \approx 1 - \f{0.6097}{n} + \f{0.05463}{n^2} \text{,}
\end{equation}
\begin{equation}
    b_n \approx 2n - \f{1}{3} + \f{4}{405n} + \f{46}{25515n^2} \text{.}
\end{equation}
$R_e$ is 2D-half-light radius, and $n$ is S\'{e}rsic index.
We set the parameters $(R_e,\, n) = (1.25\kpc,\, 1.0)$, which are typical values for $M_*=2\powten{8}\msun$ 
\citep{MassSize_Wel2014}, although the mass--size relation has a large scatter.

% Satellite DM
The satellite DM mass is $M_{\DM,\mathrm{h}} = 1\powten{11}\msun$, determined by stellar-to-halo mass relation 
\citep{SHMR}.
As shown in \citet{Ogiya2018}, the satellite DM density profile must have a flat central core. 
Therefore, we adopted Burkert profile \citep{Burkert} for it: 
\begin{equation}
    \rho(r) = 
    \f{\rho_s}{\left(1+\f{r}{r_s}\right)\left(1+\f{r^2}{r_s^2}\right)},
\end{equation}
with $(\rho_s,\, r_s) = (5.6\powten{7}\msun/\mathrm{kpc}^3,\, 4.6\kpc)$.

% Satellite GCs
We implemented 10 GCs as 10 tracer particles because \citet{Dutta_Chowdhury2020} showed that whether the GCs in DF2 are single particles or live particles does not affect their orbits so much.
We used the S\'{e}rsic profile with $(R_e,\, n) = (2.0\kpc,\, 0.5)$ as the probability distribution function for the initial positions of the 10 GCs.
Then they were placed in the range of $1 \endash 4\kpc$.
Each GC has a mass of $m_\GC = 9\powten{5}\msun$, roughly in line with those of DF2 \citep{vanDokkum2018b}.

All of the N-body components are generated by \textsc{magi} \citep{MAGI}.
In Figure~\ref{fig:initialDensityProfile}, we plot the distribution of all components.
%
% Orbit
The orbit was determined by the orbital parameters $(x_c,\, \epsilon)$ defined in 
\citet{Lacey&Cole1993}: 
\begin{eqnarray}
  x_c &\equiv& \frac{r_c(E)}{r_{200,\,\mathrm{host}}}, \\
  \epsilon &\equiv& \frac{L}{L_c(E)}, 
\end{eqnarray}
where $r_c(E)$ is the circular radius corresponding to the energy $E$, and $r_{200,\,\mathrm{host}}$ is the virial radius of the host halo.
$L$ is the satellite orbital angular momentum and $L_c(E)$ is that of the energy $E$ on a circular orbit.
The parameter $\epsilon$ is the circularity parameter, which indicates the orbit's radial nature. 
The satellite galaxy initially has $(x_c,\, \epsilon) = (0.8, 0.45)$, 
which is very similar to that adopted in \citetalias{Ogiya_main}. The slight difference in these parameter values from \citetalias{Ogiya_main} who used $(x_c,\, \epsilon) = (1.0, 0.3)$, is due to small differences in the initial conditions and simulation setup. However, our parameter values are close enough that the overall results are very similar to each other. 
With these parameters and the initial position of the satellite at $r=305$\,kpc, the initial orbital velocity of the satellite is 241\,km\,s$^{-1}$.
With the total satellite mass being $1\times 10^{11}\msun$,  the velocity dispersion of the satellite stars is about 200\,km\,s$^{-1}$.  These parameter values become important later when we try to understand the dynamics in the phase space of energy and angular momentum. 
The simulation parameters are summarised in Table~\ref{tab:params} for convenience.

\begin{table}
 \caption{Simulation Parameters}
 \label{tab:params}
 \centering
  \begin{tabular}{cc}
  \hline
  \hline
  Host halo &  \\
  \hline
    $M_{200}$ & $7.00\times 10^{12}\msun$ \\
  total mass  & $8.67\times 10^{12}\msun$ \\
  $N_{\rm DM}$ (\runone) & $1.5 \times 10^7$ \\
  $m_{\rm DM}$ (\runone) & $5.78\times 10^5\msun$ \\
  \hline\hline
    Satellite galaxy &  \\
    \hline
    total DM mass & $1.00\times 10^{11}\msun$ \\
    total baryon mass (stars+GCs)  & $2.09\times 10^{8}\msun$ \\
    $N_{\rm DM}$ & $2.5\times 10^7$ \\
    $N_{\star}$ & 49,990 \\
    $m_{\rm DM}$ & $4.00\times 10^3\msun$ \\
    $m_{\star}$ & $4.00\times 10^3\msun$ \\
 \hline\hline
    Globular clusters & \\
    \hline
    $m_{\rm GC}$ & $9.00\times 10^5\msun$ \\
    $N_{\rm GC}$ & 10 \\
    \hline 
\end{tabular}
\end{table}

%\subsection{Numerical simulation settings} 
%\label{sec:numericalSettings}

\section{Results} \label{sec:results}

\subsection{Mock observations}

We compare the evolution of the physical state of the satellite galaxy in {\runzero} and \runone.
At the same time, we also perform mock observations of our simulations and compare them to the actual DF2 observations as follows. 
%should compare our results to DF2 properties from observation.
%So we did mock observation to analyze our simulations and make plots.
%The operation is as follows.
%We define a line-of-sight to observe the satellite.
%Consider the projection of the stellar particles onto a plane perpendicular to the line of sight.
To evaluate the uncertainties introduced by the viewing angle, we shoot 13 lines of sight (LOS) that are each separated by $\pi/4$ in both longitude and latitude, and project stellar particle distribution onto the plane perpendicular to the LOS. 
We divide the projected plane into $540~\mathrm{pc}\times 540~\mathrm{pc}$ pixels.
%If a pixel has more than a certain number of stellar particles, the stellar particles within the pixel are taken into account for the analyses.
Because each stellar particle has a mass of $4000\msun$, %\gg 1\msun$, 
one particle in the above pixel roughly corresponds to the surface brightness limit of $\mu \lesssim 29~\mathrm{mag/arcsec}^2$ (assuming a mass-to-light ratio of $M/L_V \sim 3$ from \citet{Bruzual03}), which is nearly equal to the observational limit of the Dragonfly Telephoto Array \citep{Dragonfly},  
one of the telescopes that observed DF2 \citep{DF2_Nature}.
Therefore, we detect the pixels having more than one stellar particle as passing the above surface brightness limit. 
Some stellar particles are more than $10\kpc$ away from the center of the satellite galaxy due to tidal stripping, and we neglect such particles in sparse regions. 

%Figure~\ref{fig:massEvolution}, \hyperref[fig:stellarEvolution]{4} and \hyperref[fig:gcEvolution]{5} show the mean values by darker-colored lines and the minimum--maximum value by light-colored vertical lines, gained by the 13 lines of sight.
%The neighboring vertical lines are overlapped each other, so they don't look so much lines as a shade.

\subsection{Evolution of the satellite galaxy}
\label{sec:sat}

\begin{figure*}
  \centering
 \includegraphics[keepaspectratio, width=1.95\columnwidth]{fig3/Fig2.pdf}
  \caption{Projected density of satellite galaxy particles in {\runone} at $t = 0.0,\, 2.3,\, 4.9,\, 6.6,\, 7.4,\, 9.0\Gyr$ from top left to bottom right.
        The grey scale shows the DM component in the larger panels, the orange dots show the stellar component, and the blue dots indicate the positions of the GCs in the smaller inset.
        The larger panels are centered at the main host halo. 
        The inset panels are centered at the center of the satellite galaxy, and their size is about $20\kpc$. 
      }
  \label{fig:snapshots_all}
\end{figure*}

Figure~\ref{fig:snapshots_all} shows the projected density of the satellite particles (dark matter in grey color in larger panels, and star particles in orange points in the smaller inset) in {\runone} at $t = 0.0,\, 2.3,\, 4.9,\, 6.6,\, 7.4,\, 9.0\Gyr$ (the same figure for {\runzero} is shown in Appendix~\ref{app:snapshots_run0}).
All components initially have a spherical shape, so the satellite and its orbit are in the $x-y$ plane throughout the entire simulation time.
The DM particles are tidally stripped and the density around the satellite becomes as low as that of the tidal tail after $\approx 8\Gyr$.
On the other hand, the stripping of the stellar body is slower; it just expands until the fourth pericentric passage, and then a tidal tail appears after the fifth one ($\approx 6.8\Gyr$).
It keeps its spherical shape until $\approx 9\Gyr$, and after that, the satellite galaxy is completely destroyed.
In parallel with the stars, the GCs are also scattered.
We can see all the ten GCs within the small box at $t = 6.6\Gyr$ (middle right panel).

%%%%%%%%%%%%%%%%%%
%
\begin{figure*}
  \centering
     \includegraphics[keepaspectratio, width=1.95\columnwidth]{fig3/Fig3.pdf}
  \caption{Similar to Fig.~\ref{fig:snapshots_all}, but only for stars (orange dots) and GCs (blue dots).  A small square is placed at the center of mass of the satellite stellar distribution. 
        Note that each row has a different color bar from that of  Fig.~\ref{fig:snapshots_all}.
        The grey line shows the orbit of the satellite galaxy.
        The black and grey cross marks are placed at the host halo center and the initial position of the satellite galaxy, respectively.
  }
  \label{fig:snapshots_star}
\end{figure*}

Figure~\ref{fig:snapshots_star} shows the projected density of stars (orange dots) and the orbit of the satellite galaxy, together with the GC distribution (blue dots). 
We can see the entire tidal tail of stars. 
After the fourth pericenter, a sparse tidal tail appears (middle right panel).
After the fifth pericenter, it becomes denser, and finally, the stellar particles become a band-like shape.
The lack of symmetry in the tidal tail is due to the projection effect. 
We note the striking similarity between the tidal tail feature in the bottom right panel (9\,Gyr) and the observed giant tidal tail of F8D1 \citep{Zemaitis23}.
It is noteworthy that this particular long tidal feature only arises when dynamical friction is included within our simulation time, and also indicates a unique time period for the system in which the DM content should be low. 
See Appendix~\ref{app:SB} for more discussion on the surface brightness distribution.

%\color{black}
%%%%%%%%%%%%%%%%%%
%
\begin{figure}
  \centering
 \includegraphics[keepaspectratio,width=0.9\columnwidth]{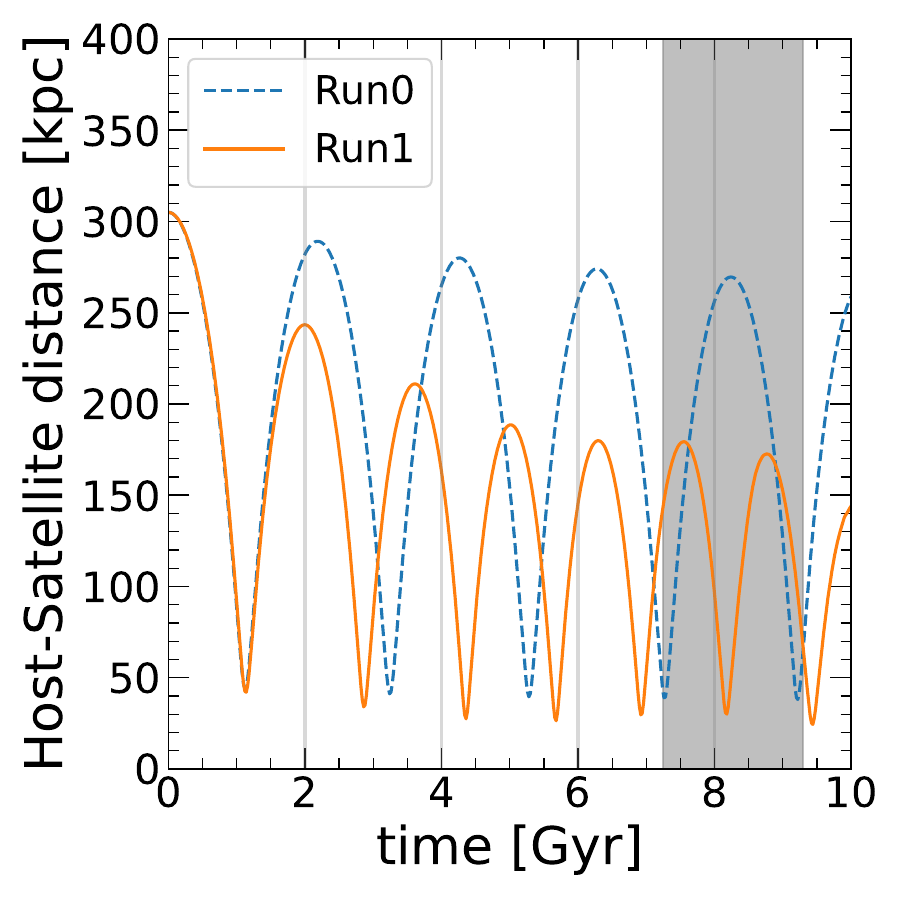}
  \caption{The distance between the centers of the host and the satellite as a function of time.
        The blue dashed and orange solid lines correspond to {\runzero} and \runone, respectively.
        We can see that the orbit in {\runone} decays more rapidly due to dynamical friction from the live dark matter halo.
        Grey shade represents the `dark matter-deficient period' (defined in the context of Figure~\ref{fig:massEvolution}, see text).
  }
  \label{fig:orbitEvolution}
\end{figure}
Figure~\ref{fig:orbitEvolution} shows the orbits of the satellite galaxies.
The ordinate shows the distance between the center of the satellite and the host.
In {\runzero}, the apocenteric distance decreases slowly,  approximately from $300\kpc$ to $250\kpc$ in $10\Gyr$\!
\!\footnote{The orbital decay observed in {\runzero} is due to self friction
\citep{Fujii2006, Fellhauer2007, Miller2020}}.
On the other hand, {\runone} shows a much faster drop, to finally $150\kpc$ in the same time span.
After some orbits, tidal interaction reduces the satellite's mass and weakens the strength of the dynamical friction, slowing the orbital decay.
The pericentric distance also shows a similar trend, declining by $6\kpc$ in {\runzero}, and by $18\kpc$ in {\runone}.
As a result, the orbital period in {\runone} is shorter than in {\runzero}, being about $2\Gyr$ in {\runzero} on average and $1.5\Gyr$ in \runone.
The number of pericentric passages is five in {\runzero} and seven in \runone, which means that the satellite in {\runone} suffered more from tidal interactions than in {\runzero}.

%%%%%%%%%%%%%%%%
%
\begin{figure}
  \centering
 \includegraphics[keepaspectratio,width=0.95\columnwidth]{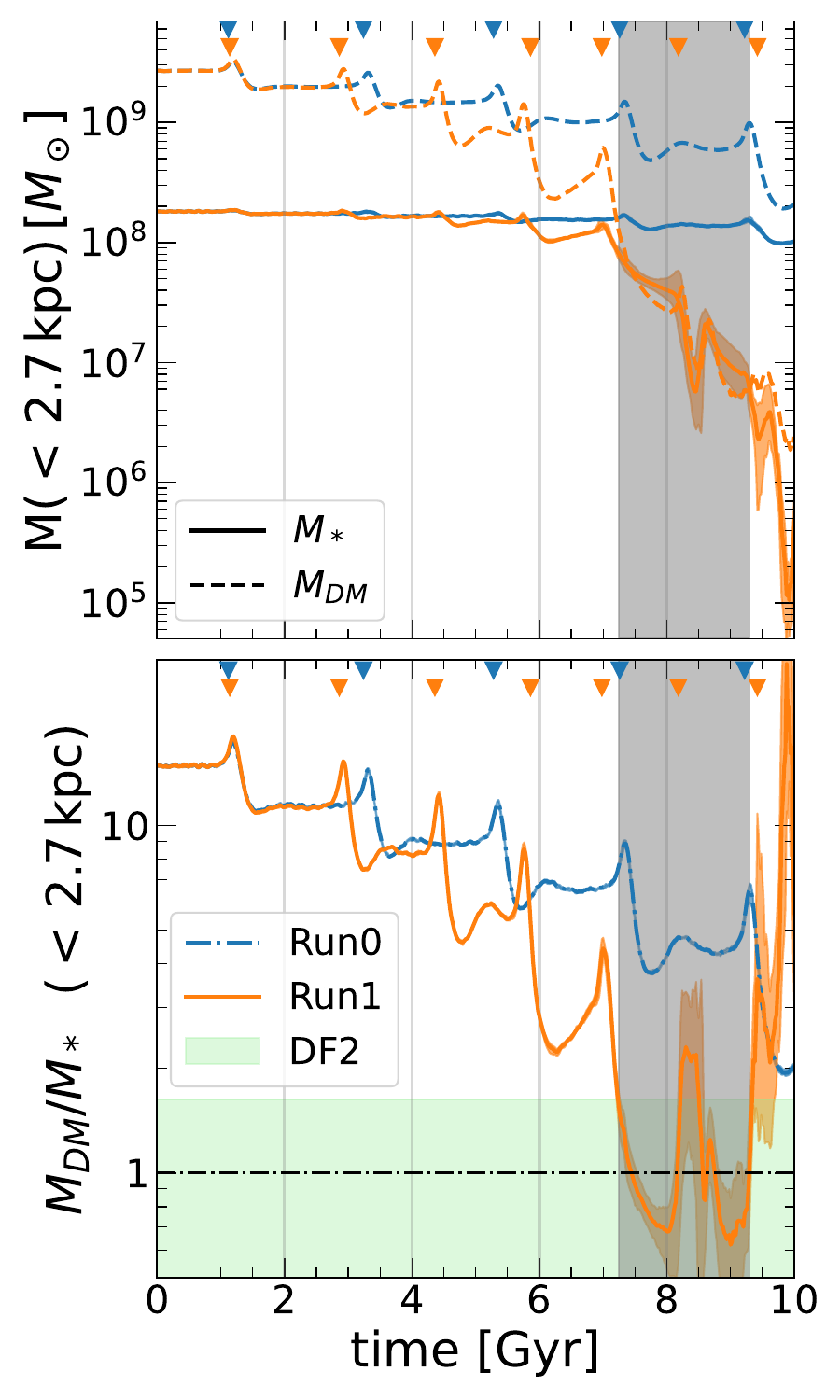}
  \caption{\textit{Top}:~Stellar (solid line) and DM (dashed line) mass within $2.7\kpc$ from the center of the satellite as a function of time.
        As for the stellar mass, the solid line and the shade around it represent the mean value and min-max range gained by 13 lines of sight, respectively.
        The triangles indicate the timing of the pericentric passages in {\runzero} (blue) and in {\runone} (orange).
        The color coding is the same as Fig.~\ref{fig:orbitEvolution}.
        \textit{Bottom}:~Mass ratio $M_\DM/M_*(<2.7\kpc)$ calculated from the top panel.
        Light-green shade represents the observed range of DF2 \citep{Danieli2019}.
        We defined the `dark matter-deficient period', shown by the grey shade, which largely overlaps with the green shade.
  }
  \label{fig:massEvolution}
\end{figure}

\begin{figure}
  \centering
    \includegraphics[keepaspectratio,width=\columnwidth]{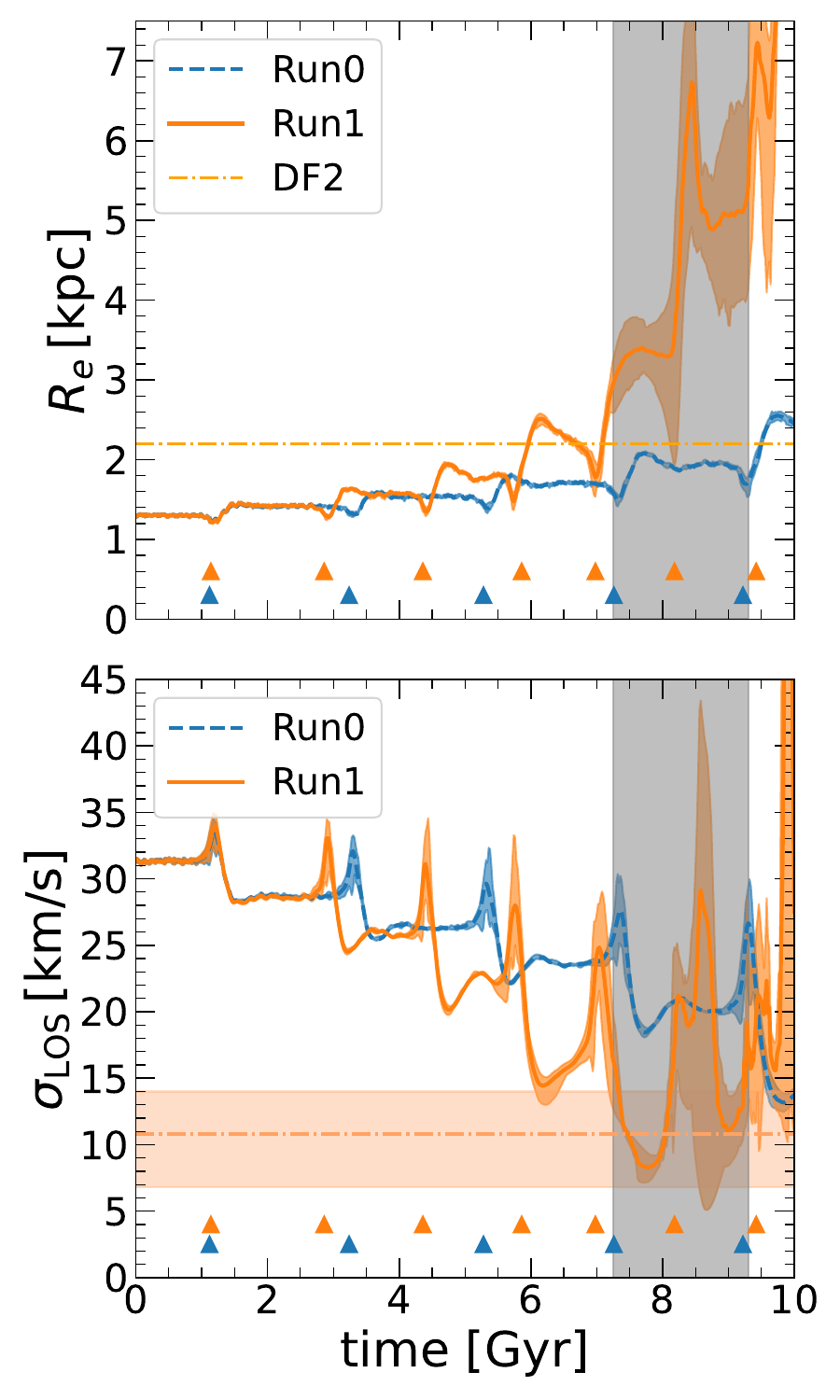}
  \caption{%The makeup is the same as Figure~\ref{fig:massEvolution}.
        \textit{Top}:~2D half-light radius $R_e$.
        The measured value of DF2 is indicated by a horizontal dashed line \citep{DF2_Nature}.
        \textit{Bottom}:~Velocity dispersion of the satellite's stars along line of sight. 
        The horizontal line and shade denote the DF2 value \citep{Emsellem2019}.
        %Both properties are consistent with DF2 at the same time as ($\approx 7.3\endash {\bf 9.3}\Gyr$), {\bf except for the short period of pericentric passage around 8.2\Gyr.} 
        Grey shade represents the `dark matter-deficient period' (see the main text). The triangles indicate the timing of the pericenteric passages in {\runzero} (blue) and in {\runone} (orange).
        }
  \label{fig:stellarEvolution}
\end{figure}

\begin{figure}
  \centering
   \includegraphics[keepaspectratio,width=0.9\columnwidth]{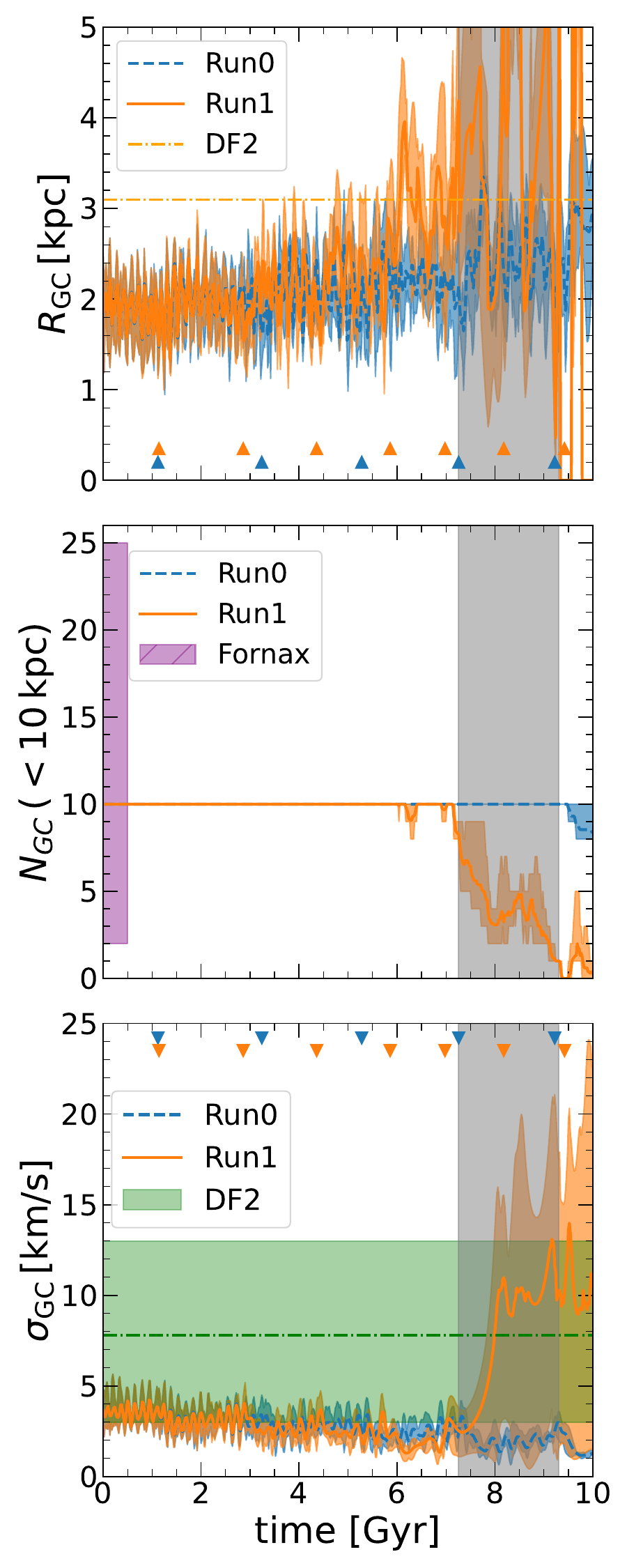}
  \caption{Similar setup as Figs.~\ref{fig:massEvolution} and \ref{fig:stellarEvolution}.
        \textit{Top}: 2D projected median radius ($\Rgc$) of the ten GC particles. 
        This has a similar trend to the stellar half-light radius, but the time when it reaches the DF2 value is slightly earlier than the dark matter-deficient period.
        \textit{Middle}: Number of GCs within projected radius of $10\kpc$ from the satellite centre. It begins to decrease rapidly during the dark matter-deficient period. Purple hatch on the left edge indicates observational $1\sigma$ range of low surface brightness galaxies in Fornax cluster \citep{Prole2019}.
        \textit{Bottom}: Line-of-sight velocity dispersion of GCs, which shows a rapid increase in the dark matter-deficient period. The green shade indicates the observational likelihood estimate of $\sigma_{\rm GC} = 7.8^{+5.2}_{-2.2}$\,km\,s$^{-1}$ by \citet{Dokkum18}. 
  }
  \label{fig:gcEvolution}
\end{figure}

Figure~\ref{fig:massEvolution} shows the evolution of the mass and mass ratio of the satellite system.
The upper panel shows the masses of stars and DM ($M_*$ and $M_\DM$) within $2.7\kpc$ around the centre of the satellite as a function of time.
The lower panel plots their mass ratio: $M_\DM/M_* (<2.7\kpc)$.
When this mass ratio approaches unity closely, the satellite galaxy is regarded as a DMDG.
In Figures~\ref{fig:massEvolution}, \ref{fig:stellarEvolution} and \ref{fig:gcEvolution}, the mean values of 13 LOS results are shown by darker-colored lines, and the min--max range of 13 LOS is shown by light-colored shade around them. 

Tidal stripping reduces the galaxy mass in an outside-in fashion.
Because the stellar body is more centrally concentrated than the DM halo, the stellar mass is almost constant until the fourth pericentric passage, whereas the DM mass begins to decrease from the first pericentric passage.
Therefore, the DM mass will be less than the stellar mass at some point.
In the case of {\runzero}, it does not happen within $10\Gyr$, and the satellite galaxy could not become a DMDG within the simulation time.
On the other hand, in \runone, the satellite became DMDG at $\approx 7.3\Gyr$, just after the fifth pericentric passage, as shown in Fig.~\ref{fig:massEvolution}.

After the sixth pericenteric passage, the dark matter-deficient satellite galaxy is completely destroyed at $\approx 8.6\Gyr$; it cannot maintain its spherical shape and becomes a band-like shape. 
The satellite galaxy becomes largely dark matter-deficient during $7.3\endash 9.3\Gyr$,  except for a brief period near the sixth pericentric passage.
This `\textit{dark matter-deficient period}' is shown by the grey-shaded region in the figures, where the DM fraction largely overlaps with that of DF2. 
Due to tidal heating, a tidally born DMDG has a very small binding energy, which can be easily dissipated by the next tidal interaction, at which point its stellar distribution cannot be defined well relative to the dark matter distribution.

%%%%%%%%%%%%%%%%%%%%%%%%%%%
%

Figure~\ref{fig:stellarEvolution} shows the evolution of the stellar component of the satellite.
The upper panel plots the 2D-projected half-light radius as a function of time, which we approximated by the half-stellar-mass radius on the projected plane. 
We can see that the stellar body expands right after each pericentric passage.
This is because the galaxy gains energy during the high-speed passage (tidal heating), and its effect is much stronger in \runone. 
This half-light radius in {\runzero} did not reach the DF2 value, but it already goes over the DF2 value in {\runone} during the dark matter-deficient period.
The lower panel plots the line-of-sight velocity dispersion of the observable stellar particles.
Reflecting the expansion, it decreases as the half-light radius increases.
This is also consistent with the DF2 value during the DM-deficient period.
In the bottom panel of Fig.~\ref{fig:stellarEvolution}, we see that the velocity dispersion along the line of sight, $\sigma_{\rm LOS}$, becomes consistent with that of DF2 at $t\sim 7.3-8.2$\,Gyr, but followed by a sharp increase after the sixth pericentric passage. 

So far, we have seen that the satellite's stellar properties become roughly consistent with DF2 during the DMDG phase. 
Finally, let us examine the GC distribution. 

%%%%%%%%%%%%%%%%%%%%%%%%%%%%%%%
%

\subsection{Distribution of GCs}

Figure~\ref{fig:gcEvolution} shows the evolution of the GC distribution.
The upper panel plots the 2D-projected median radius of GCs ($\Rgc$) as a function of time.
%\color{red}
It does not include the GCs farther away than $10\kpc$ nor stellar particles because they are regarded stripped.
%\color{black}
Although this definition is basically the same as that of stars, $\Rgc$ shows a different evolution from $R_e$.
We find that $\Rgc$ oscillates rapidly, which is due to their random motion within the satellite galaxy. %could be due to the random scattering of GC particles within the satellite galaxy. 
%This is actually seen in the stellar half-light radius, but $\Rgc$ has a much higher amplitude, just because of the small number of the GCs.
%It is interesting that its increase is much faster than that of the stars shown in Figure~\ref{fig:massEvolution}.
In \runone, $\Rgc$ reaches the DF2 value around fourth pericenter, 
%(fifth for the stellar half-light radius), 
and rapidly increases after the fifth pericenter. 
%(sixth for the stellar half-light radius).
%It seems to show a similar trend in {\runzero}, too.
{\runzero} shows a similar trend, but with a much slower speed. 
The difference is not due to dynamical friction on the satellite, 
because if this were the case, we would only see this trend in \runone. 
%, but both runs have the same feature\!
%\!\footnote{
Note that the GCs suffer from dynamical friction caused by the density field of the satellite system itself. 
%}.
%This may be because of the difference in 
Initialisation of the GC distribution might also have some impact (see Section~\ref{sec:comp_ogiya} for a further discussion).

The middle panel of Figure~\ref{fig:gcEvolution} plots the number of GCs within $10\kpc$ from the satellite centre.
It decreases rapidly during the dark matter-deficient period, ranging from two to ten.
% the minimum number is two (out of ten) in the midst of the sixth pericentric passage.
This suggests that $0\endash80\%$ of GCs can be stripped, and that the candidate galaxy of DF2 should initially have approximately $10\endash50$ GCs before the majority of them are stripped \citep{DF2_Nature, Shen2021a}.
Therefore, if we put more GCs initially, we can probably reproduce the number of GCs within $10\kpc$ of DF2. 
Whether we can reproduce the exact number of GCs within a certain radius is not that critical.  Rather, statistical signatures such as the distribution in the phase space would be more interesting as we will discuss in Sec~\ref{sec:phase}. 

In the bottom panel of Figure~\ref{fig:gcEvolution}, we show the line-of-sight velocity dispersion of GCs ($\sigma_{\rm GC}$).  Both runs show a gradual decrease in $\sigma_{\rm GC}$, consistent with previous work \citep{Ogiya2022,Zhang24}.
This is because GCs are subject to the dynamical friction within the satellite galaxy, which causes them to sink gradually towards the centre of the satellite galaxy. 
However, interestingly, {\runone} shows a sharp increase just after entering the DM-deficient period, whereas {\runzero} continues to show a decreasing trend. This is because the tidal effect becomes stronger sooner in {\runone} than in {\runzero} due to faster orbital decay by dynamical friction with closer pericentre passages.  Once the satellite galaxy is completely disrupted tidally, then $\sigma_{\rm GC}$ stabilises to about 10\,km\,s$^{-1}$ with some fluctuations.

\section{Effect of Dynamical friction} \label{sec:dynamicalFriction}

We see that dynamical friction can turn a galaxy into a DMDG in a short time.
The fundamental effect of dynamical friction is to extract momentum from a moving object.
In our case, the satellite galaxy loses its momentum, leading to orbital decay.
The amount of mass loss due to tidal stripping depends mainly on the pericentric distance
\citep{Jackson2021, Montero-Dorta2022, Smith2022}, so the orbital decay may contribute to the enhancement of the tidal stripping.
However, when comparing the pericentric distance in {\runzero} and \runone, the difference is not so large (Figure~\ref{fig:orbitEvolution}): its ratio is at most $1.47$ at the fourth pericenter.
Therefore, the difference in the changes of the satellite properties is small (e.g., the decrease in the mass ratio $\Delta (M_\DM/M_*)$ after the second pericentric passage is almost identical in the two runs as seen in Fig.~\ref{fig:massEvolution}).
Thus, the dynamical friction does not enhance the tidal effect so much, but shortens the orbital period and increases the number of pericentric passages (tidal interaction) per unit time.

% \subsection{Comparison with the literature}
\section{Comparison with the literature} 
\label{sec:comp_ogiya}

We simulated almost the same configuration as \citetalias{Ogiya_main} in {\runzero}.
Note that \citetalias{Ogiya_main} used a time-growing potential, although we used a static one.
The growing host halo results in orbital decay of its satellite galaxies over the first few orbits \citepalias{Ogiya_main}, and the evolutionary timescale is shorter than that of {\runzero} and longer than that of \runone.
The satellite in \citetalias{Ogiya_main} became a DMDG in $\sim$$9\Gyr$, which is the midpoint of {\runzero} and \runone.

\citetalias{Ogiya_main} succeeded in reproducing the GC distribution of DF2 when the satellite galaxy is a DMDG.
In our simulation, the GC distribution expanded a little too fast.
Comparing our Fig.~\ref{fig:gcEvolution} with Fig.~5 in \citetalias{Ogiya_main} which plotted the same physical quantity, the only clear difference is the higher amplitude of rapid oscillation in our case.
We initially place the ten GCs using S\'{e}rsic profile as a probability distribution function. 
\citetalias{Ogiya_main} adopted a different method, where they first put stellar particles and transformed ten of those located at $r\approx 2.5\kpc$ into GCs.
After $\sim\,100\Myr$, their distribution is well fitted by the S\'{e}rsic profile that we use ($R_e=2.0\kpc,\, n=0.5$).
These slight differences in implementation could easily lead to a somewhat different evolution of their distribution, which is not surprising. 

Purple hatch on the bottom panel of Fig.~\ref{fig:gcEvolution} shows $1\sigma$ range of fitted number of GCs in low surface brightness galaxies (LSBGs) in the Fornax cluster of absolute magnitude $M_V \approx -15.4$\!
\!~\footnote{
    The absolute magnitude of DF2 is $M_V =-15.4$ \citep{DF2_Nature}.
}
\citep{Prole2019}. % g-band
LSBGs of similar stellar mass observed in the local universe have $\sim 0-10$ GCs; 
its range is wide and its measurement error is also large \citep{Forbes2020, Jones2023}.  Therefore, putting more GCs in our initial condition is a reasonable remedy if we want to try to make our simulation agree more with DF2 observations.
In any case, the present observational precision of GC distribution at high redshift is insufficient to distinguish these differences.

\section{Phase space evolution} 
\label{sec:phase}

Examining the distribution of stars and GCs in the phase space of total energy $E_{\rm tot}$ versus $z$-direction angular momentum $L_z$ enhances our understanding of kinematical signatures of DMDGs better. 
In Fig.~\ref{fig:phase}, we show the time evolution of phase space at $t=0.0, 2.9, 4.7, 5.2, 7.5, \&\, 9.5$\,Gyr. The background grayscale shows the dark matter mass per pixel in phase space.  The central rectangle shows the range of values that local Milky Way stellar streams are found \citep{Bonaca21}. The comparison indicates that the satellite galaxy in our simulation setup has a higher angular momentum in a deeper potential well than most of the Milky Way stellar streams. 

At $t=0$, the satellite stars (cyan points) are distributed in a narrow horizontal strip with roughly constant energy but exhibit a wide distribution in $L_z$ due to the internal velocity dispersion of the satellite galaxy. 
The satellite experiences its second periastron at $t=2.9$\,Gyr, where the cyan stellar points show a narrow vertical distribution due to the deeper potential well at pericentre (stretching the distribution downward) and the effects of tidal heating (pulling the distribution upward). 
At $t=4.7$\,Gyr, the satellite is in an intermediate state following the third periastron and occupies a localized region of phase space determined by its initial orbital conditions. At $t=5.2$\,Gyr, the satellite moves through the apocenter after the third periastron, reverting to a similar distribution as the initial state at $t=0$\,Gyr, although with a somewhat shorter horizontal extension due to gradual energy loss from dynamical friction. 
At $t=7.5$\,Gyr, the satellite enters the DM-deficient phase, and its distribution becomes similar to that at pericentre at $t=2.9$\,Gyr. 
As observed in the accompanying movie, the phase space distribution of stars and GCs appears roughly ``frozen" in phase space after $t\sim 8$\,Gyr, with the entire distribution becoming somewhat narrower vertically from $t=7.5$ to $9.5$\,Gyr. This is evident in the final state of the DM-deficient phase at $t=9.5$\,Gyr in the bottom right panel. 

In summary, our phase space analysis shows that a narrow vertical distribution is a very good indicator of pericentre passage. 
A movie that summarises the orbital and phase space evolution of our satellite galaxy can be seen at the following URL as well as the journal website: \url{https://www.youtube.com/shorts/_XJtq6uy_EY}

\begin{figure*}
  \centering
 \includegraphics[keepaspectratio, width=0.95\columnwidth]{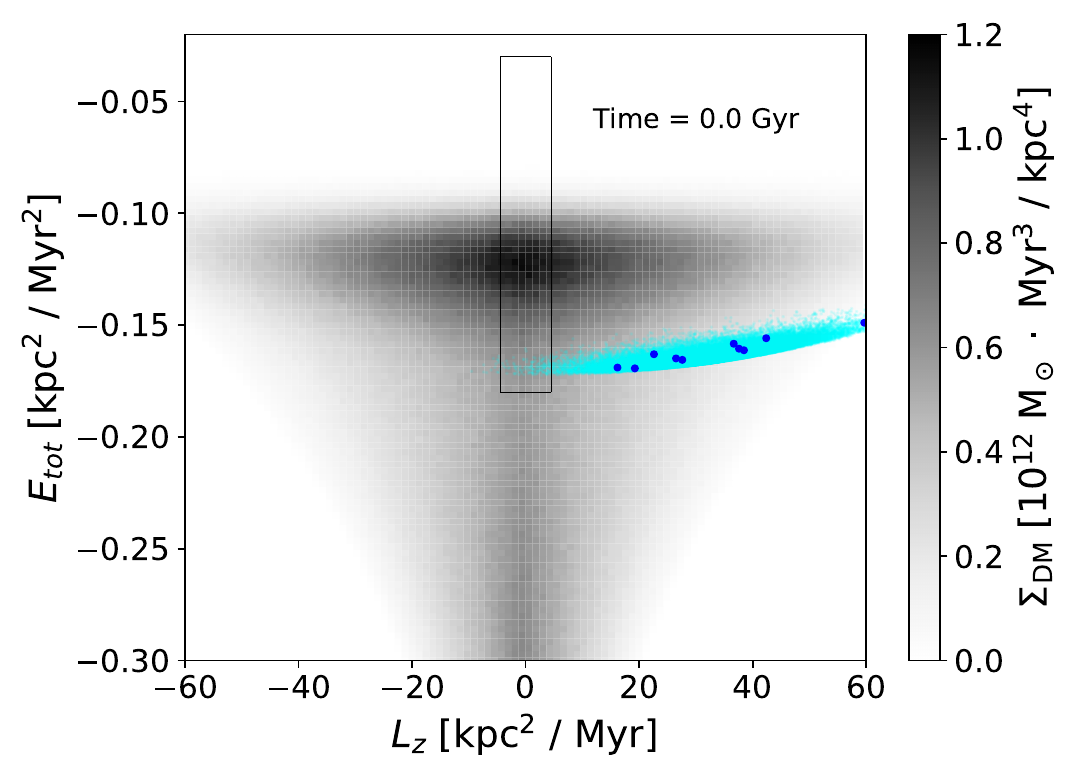}
  \includegraphics[keepaspectratio, width=0.95\columnwidth]{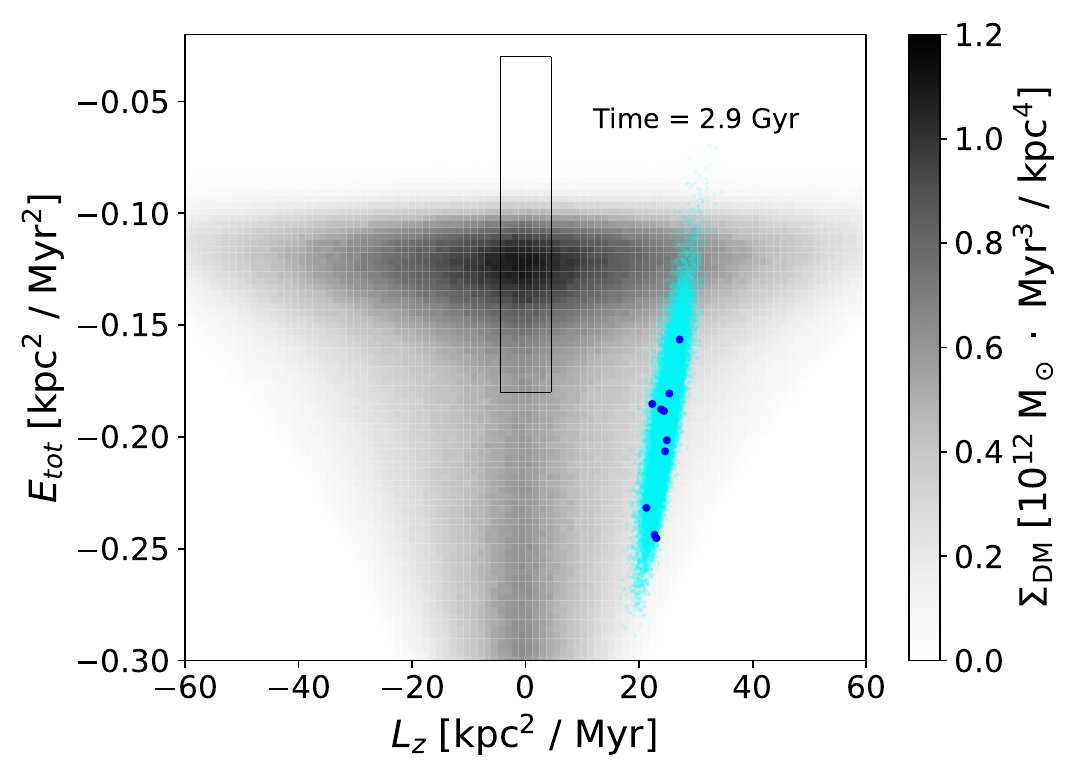}\\
 \includegraphics[keepaspectratio, width=0.95\columnwidth]{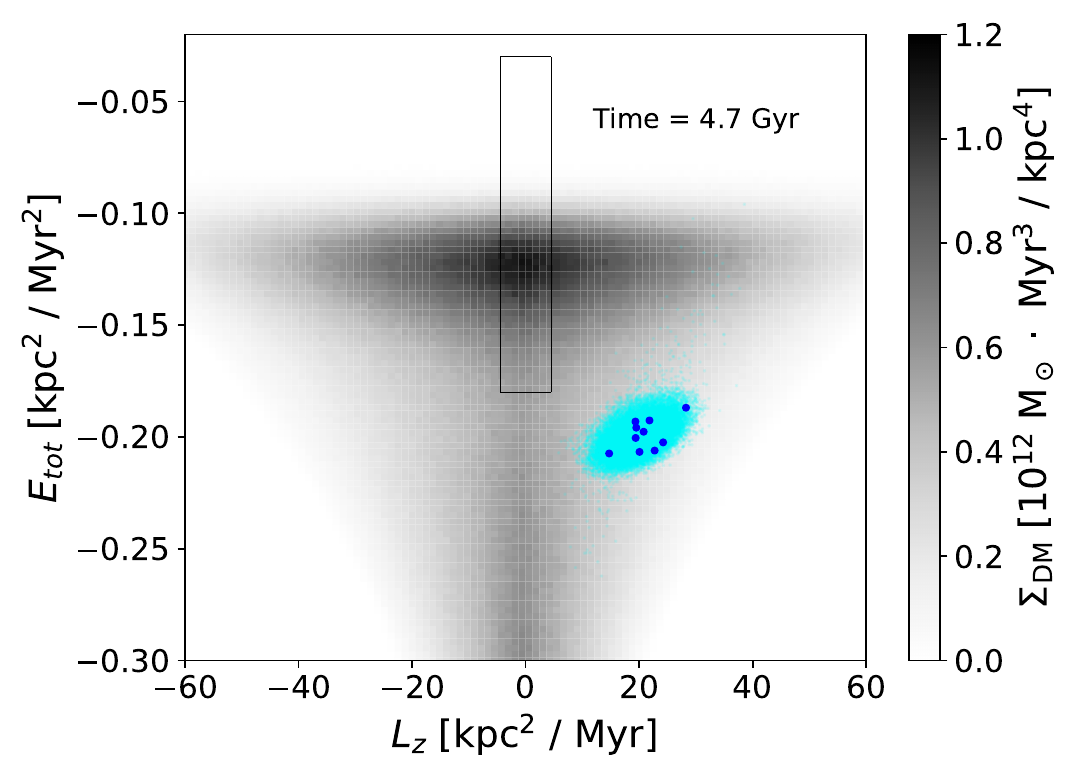}
  \includegraphics[keepaspectratio, width=0.95\columnwidth]{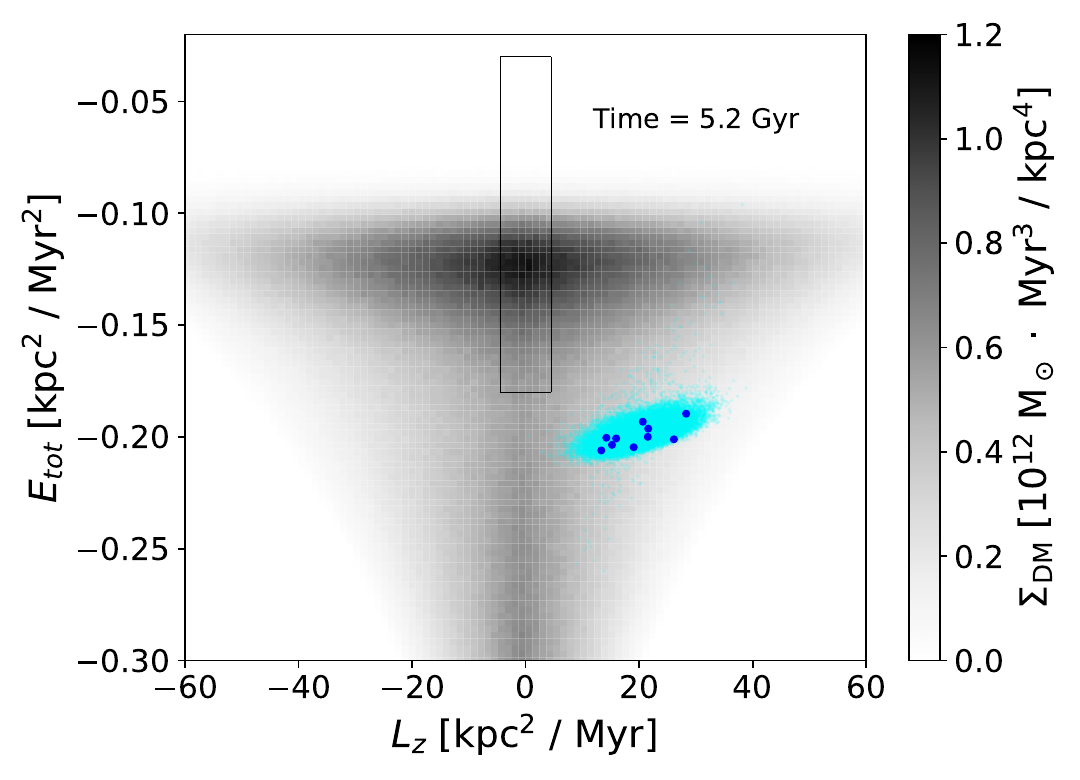}\\
  \includegraphics[keepaspectratio,
    width=0.95\columnwidth]{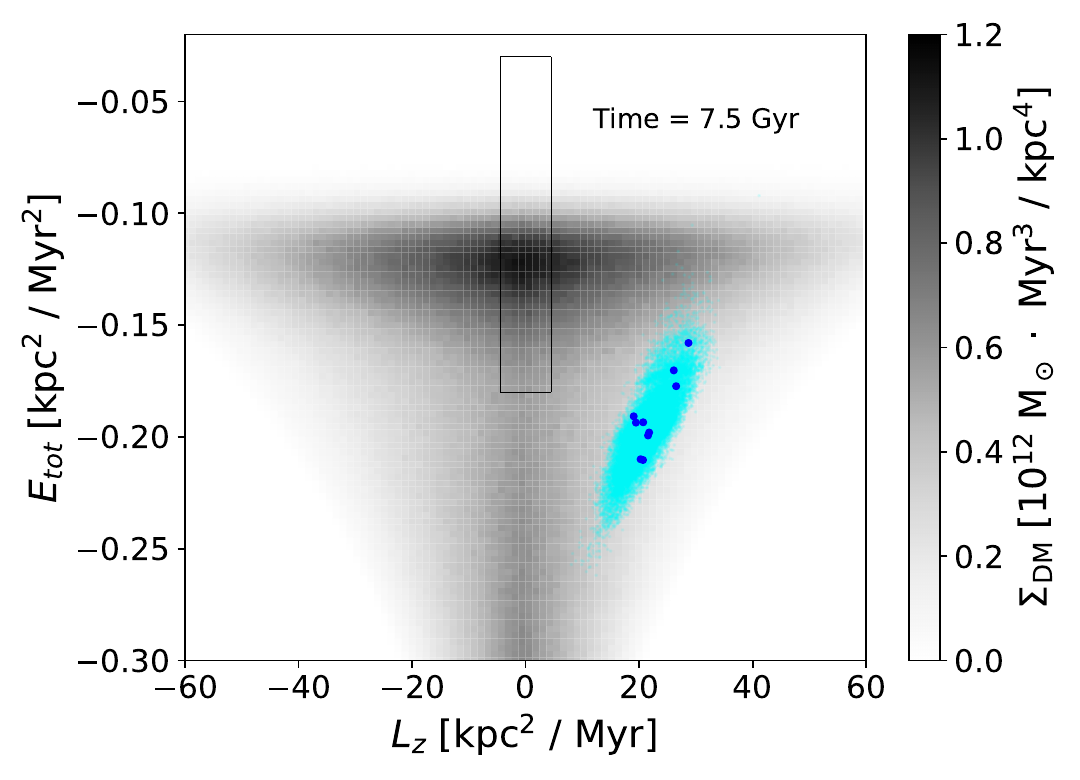}
  \includegraphics[keepaspectratio, width=0.95\columnwidth]{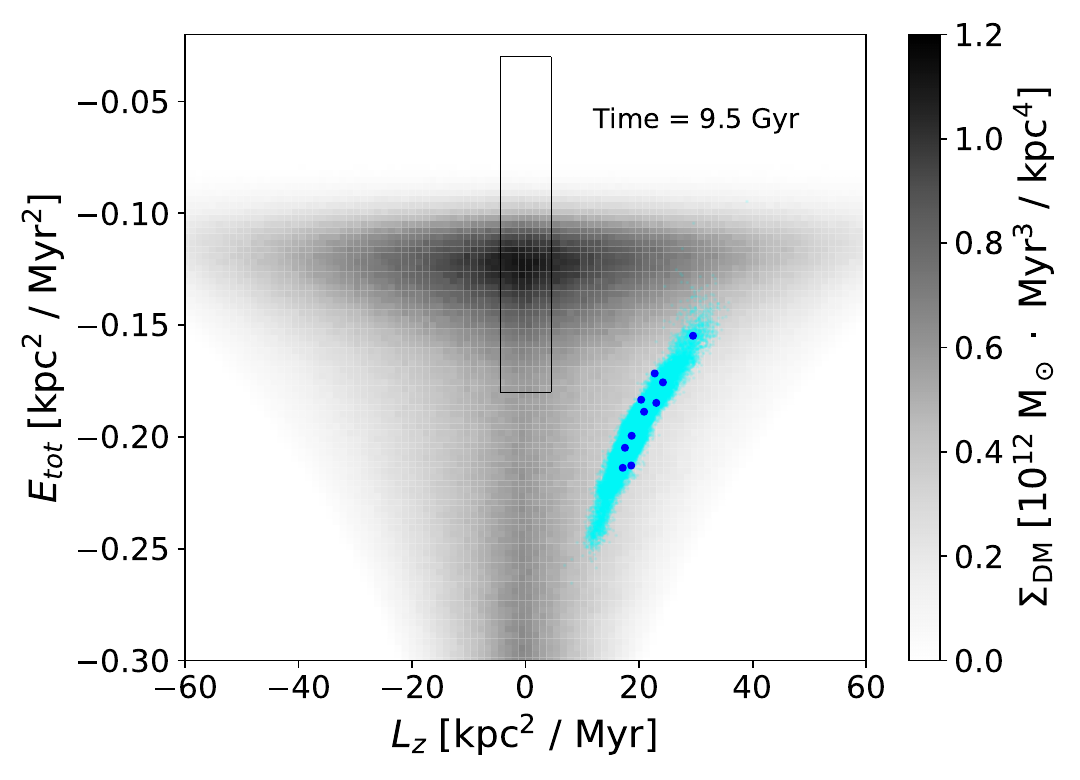}  
  \caption{Phase space evolution of the stars of satellite galaxy is shown by cyan points at $t=0.0$\,Gyr (initial state), $t=2.9$\,Gyr (second periastron), $t=4.7$\,Gyr (intermediate state after third periastron), $t=5.2$\,Gyr (apocenter after third periastron), $t=7.5$ (entering the DM-deficient phase), \& $t=9.5$\,Gyr (just after the DM-deficient phase). 
  Darker blue points are ten GC particles. Background grayscale is the dark matter mass in each pixel of the phase space plane. 
  }
  \label{fig:phase}
\end{figure*}

\section{Conclusion \& Discussion} \label{sec:conclusion}

Many numerical simulations of DMDGs have been performed under the assumption that dynamical friction can be ignored. 
In this work, we ran two sets of simulations to measure the effect of dynamical friction, one with a full N-body and the other with an analytical host halo. 

Our primary conclusion is that dynamical friction \textit{cannot} be ignored, as it plays a major role in producing tidal features like those observed in galaxies such as F8D1. Dynamical friction shortens the timescale for galaxies to evolve into DMDGs by a few Gyr in our setup, which mirrors the conditions that reproduce DF2 in \citetalias{Ogiya_main}. 
This finding suggests that even galaxies on more circular orbits can become DMDGs, extending the possible orbital parameter space in which DMDG formation can occur. 
Our results emphasize that the role of dynamical friction is crucial in enabling the formation of DMDGs across a broader range of orbital configurations. 
We note that the lifetime of the tidally formed DMDGs is only about one orbital period because they get more energy than the binding energy and are easily destroyed in the next tidal interaction.
As our simulations show, the satellite galaxy freezes in the phase space of total energy versus angular momentum and maintains a high-surface brightness core and extended tidal tail even at $t=10$\,Gyr (see bottom right panel of Fig.~\ref{fig:SB}). However, the measurement of $M_{\rm DM}/M_\star$ becomes more uncertain with the stripped dark matter and long tidal tails, making the definition of a DMDG less clear over time.

While these simulations demonstrate the importance of dynamical friction, it is essential to recognize that the satellite galaxy’s evolution is sensitive to initial conditions such as the mass ratio, orbital energy, and orbital eccentricity (circularity). These factors can significantly influence the extent to which dynamical friction contributes to the satellite’s transition into a DM-deficient regime.

For example, the mass ratio between the satellite and host galaxy is a key determinant of the strength of dynamical friction. In cases where the mass ratio is lower, the dynamical friction force would be weaker, leading to a slower orbital decay and fewer pericentric passages within a given time, potentially delaying the satellite’s evolution into a DMDG. Conversely, a higher mass ratio would enhance the dynamical friction effect, causing the satellite to lose orbital energy more rapidly and accelerating the stripping of dark matter. 
Thus, the efficiency of dynamical friction is closely tied to the mass ratio, with larger satellites experiencing more pronounced orbital decay and stronger tidal interactions.
 In our simulations, the chosen mass ratio is in the optimal range for producing effective orbital decay, but this sensitivity highlights the need to explore other values in future studies.

%Our choice of satellite and host masses are in the sweet spot of effective orbital decay ($M_{\rm sat} / M_{\rm host} \gtrsim 0.01$). 
%Hence, it is not surprising that we find a strong effect of dynamical friction. The observed masses of the host and the infalling satellite are both uncertain, and the time it takes to reach the DMDG regime depends on these somewhat arbitrary parameter choices.

Orbital energy of the satellite also plays a significant role in determining the system’s evolution. A higher initial orbital energy would result in wider orbits, where the satellite experiences weaker tidal forces, potentially postponing dark matter loss unless dynamical friction acts to tighten the orbit. In contrast, a satellite with lower orbital energy would naturally have a smaller orbit and more frequent close encounters with the host galaxy, allowing tidal stripping to occur even without significant dynamical friction. Therefore, the initial orbital energy sets the stage for the satellite’s susceptibility to tidal interactions, and the presence of dynamical friction can either accelerate or moderate this process depending on the initial energy level.

Orbital circularity is another critical factor influencing the formation of DMDGs. 
Satellites on highly eccentric orbits experience stronger tidal forces at pericenter, leading to enhanced mass stripping, even without significant dynamical friction. 
In such cases, the role of dynamical friction may be less pronounced, as the eccentric orbit itself generates sufficient tidal interactions to drive the system into a DM-deficient state. 
On the other hand, for satellites on more circular orbits, where pericentric passages are less frequent and less extreme, dynamical friction becomes more important in gradually reducing the satellite’s orbital radius, enabling repeated tidal interactions that lead to the depletion of dark matter over time.
This is consistent with findings from \citet{Zhang24}, who demonstrated that increasing orbital circularity weakens tidal stripping by pushing the pericenter farther from the host galaxy center.

While our simulations focus on a particular set of initial conditions, they demonstrate that dynamical friction plays a crucial role in this scenario. However, varying the mass ratio, orbital energy, or circularity could produce different evolutionary outcomes, including cases where dynamical friction is less critical or even unnecessary for DMDG formation. Future studies exploring a broader range of initial conditions could provide a more comprehensive understanding of how these factors interact and affect the tidal evolution of satellite galaxies. Despite this sensitivity, our current results provide valuable quantitative insights into the role of dynamical friction in satellite galaxy evolution and establish a framework for future investigations.

Finally, we note that the globular clusters (GCs) in our simulations spread out more rapidly than those observed in DF2, resulting in a GC distribution that does not perfectly match the observations.
However, by refining the initial number and distribution of GCs, it may be possible to better reproduce the observed GC configuration through additional iterations.
That said, the precise final positions of the GCs are not our primary concern. More importantly, their kinematical properties provide valuable insights into the underlying tidal formation mechanism of DMDGs. 
In particular, our simulations indicate that the line-of-sight velocity dispersion of GCs becomes elevated during the DM-deficient phase, which could serve as a key observational signature of this evolutionary stage (Fig.\ref{fig:gcEvolution}). 
Additionally, the orbital and phase space evolution of the satellite stars reveals a vertically narrow feature in the total energy versus angular momentum space, a clear indicator of pericenter passage (Fig.~\ref{fig:phase} and the associated movie). Given the lack of observational data on GCs at higher redshifts ($z \sim 1$), when they were likely formed, and the large scatter in the local GC distribution, our simulations provide a plausible scenario in which dynamical friction plays a central role in DMDG formation.

\section*{Acknowledgements}

We would like to thank the anonymous referee, Go Ogiya, Frank van den Bosch, Yoshiyuki Inoue, Shinsuke Takasao, Nicolas Ledos, Keita Fukushima, and Yuri Oku for various help and useful comments on our work. 
We thank Annette Ferguson for useful comments on the comparison between the F8D1 system and our work.
We are grateful to Volker Springel for providing the original version of {\sc GADGET-3}, on which the {\sc GADGET3-Osaka} code is based. 
Some of the numerical computations for this study were carried out on the Cray XC50 at the Center for Computational Astrophysics, National Astronomical Observatory of Japan, and the {\sc SQUID} at the Cybermedia Center, Osaka University as part of the HPCI system Research Project (hp230089, hp240141). 
This work is supported in part by the MEXT/JSPS KAKENHI grant number 20H00180, 22K21349, 24H00002, and 24H00241 (K.N.). 
K.N. acknowledges the support from the Kavli IPMU, World Premier Research Center Initiative (WPI), UTIAS, the University of Tokyo.

%%%%%%%%%%%%%%%%%%%%%%%%%%%%%%%%%%%%%%%%%%%%%%%%%%
\section*{Data Availability}

Numerical data is available upon request to the authors. 
 
% The inclusion of a Data Availability Statement is a requirement for articles published in MNRAS. Data Availability Statements provide a standardised format for readers to understand the availability of data underlying the research results described in the article. The statement may refer to original data generated in the course of the study or to third-party data analysed in the article. The statement should describe and provide means of access, where possible, by linking to the data or providing the required accession numbers for the relevant databases or DOIs.



\bibliographystyle{mnras}
\bibliography{lib} % if your bibtex file is called example.bib

%\begin{thebibliography}{99}
%\end{thebibliography}

%%%%%%%%%%%%%%%%%%%%%%%%%%%%%%%%%%%%%%%%%%%%%%%%%%

%%%%%%%%%%%%%%%%% APPENDICES %%%%%%%%%%%%%%%%%%%%%

\appendix

\section{Snapshots of {\runzero}} 
\label{app:snapshots_run0}

Here we present the same figures for {\runzero} as  Fig.~\ref{fig:snapshots_all} and \ref{fig:snapshots_star}.
Figure~\ref{fig:snapshots_all_run0} shows that both DM and stars maintain a spherical shape throughout the entire simulation time, because {\runzero} assumes a spherical background host-halo potential which does not apply dynamical friction on the satellite halo. 

The size of each frame in Figure~\ref{fig:snapshots_star_run0} is the same as Fig.~\ref{fig:snapshots_star} which is set to contain the entire orbit of {\runone}. Therefore the orbital track of {\runzero} goes beyond the frame. %which we can notice in Fig.~\ref{fig:orbitEvolution} as well.
The orbital track also tells us that the orbital period is longer in {\runzero} than in {\runone}.

%%%%%%%%%%%%%%%%%%
%
\begin{figure*}
  \centering
 \includegraphics[keepaspectratio, width=1.9\columnwidth]{fig3/FigA1.pdf}
  \caption{Same as Fig.~\ref{fig:snapshots_all}, but for \runzero.
        Projected density of satellite particles in {\runzero} at $t = 0.0,\, 2.3,\, 4.9,\, 6.6,\, 7.4,\, 9.0\Gyr$ from top left to bottom right.
        In the large panels, greyscales show the DM component, orange scales show the stellar component, and blue dots indicate the positions of the GCs in the small panels.
        The small panels are centered at the center of the satellite galaxy, and their frames are illustrated as black squares in the large panels.
        Their length per side is $10\kpc$.
  }
  \label{fig:snapshots_all_run0}
\end{figure*}
\begin{figure*}
  \centering
    \includegraphics[keepaspectratio, width=1.9\columnwidth]{fig3/FigA2.pdf}
  \caption{Same as Fig.~\ref{fig:snapshots_star}, but for \runzero.
        Note that each row has a different color bar, different from Fig.~\ref{fig:snapshots_all_run0}.
        The grey line shows the orbit of the satellite galaxy.
        The black and grey cross marks are placed at the centre of the host halo and at the initial position of the satellite galaxy, respectively.
  }
  \label{fig:snapshots_star_run0}
\end{figure*}

\section{Surface brightness distribution of the disrupted satellite} 
\label{app:SB}

In Sec~\ref{sec:sat}, we noted a striking similarity between the tidal tail feature in the bottom right panel of Fig.~\ref{fig:snapshots_star} at $t=9$\,Gyr and the observed giant tidal tail of F8D1 \citep[][see their Fig.\,3]{Zemaitis23}.
F8D1 is a system that has been neglected for decades despite its proximity. There is currently no information on its dark matter or GC content, however, more new data and results are expected to come soon and we can perform more rigorous comparisons with our simulations (private communication with A. Ferguson). 

\begin{figure*}
  \centering
     \includegraphics[keepaspectratio, width=0.9\columnwidth]{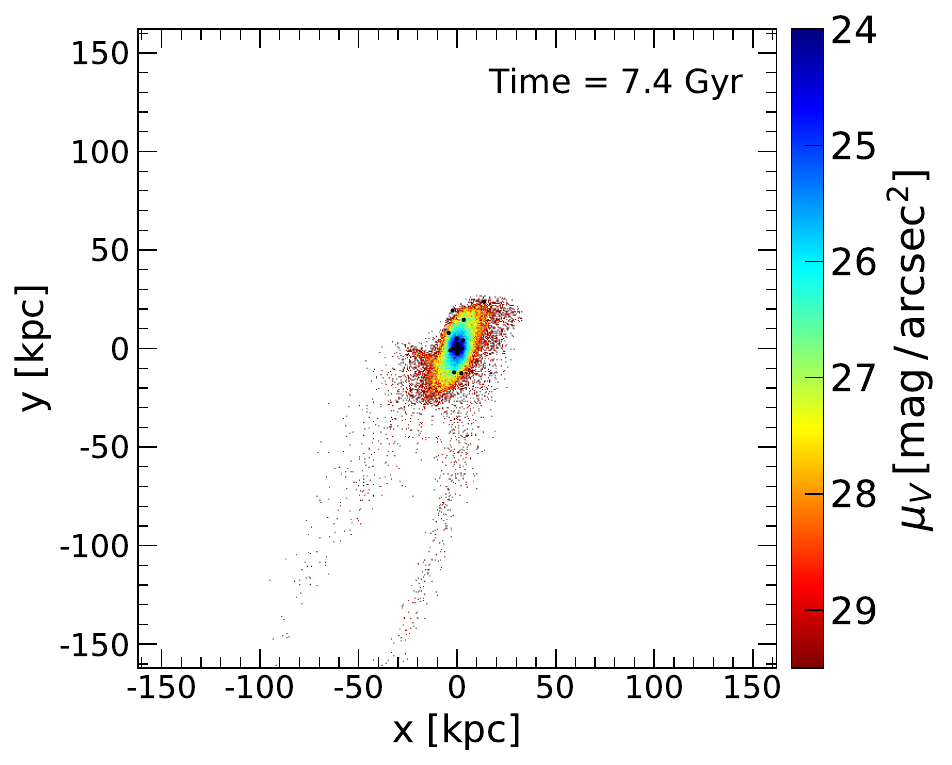}
     \includegraphics[keepaspectratio, width=0.9\columnwidth]{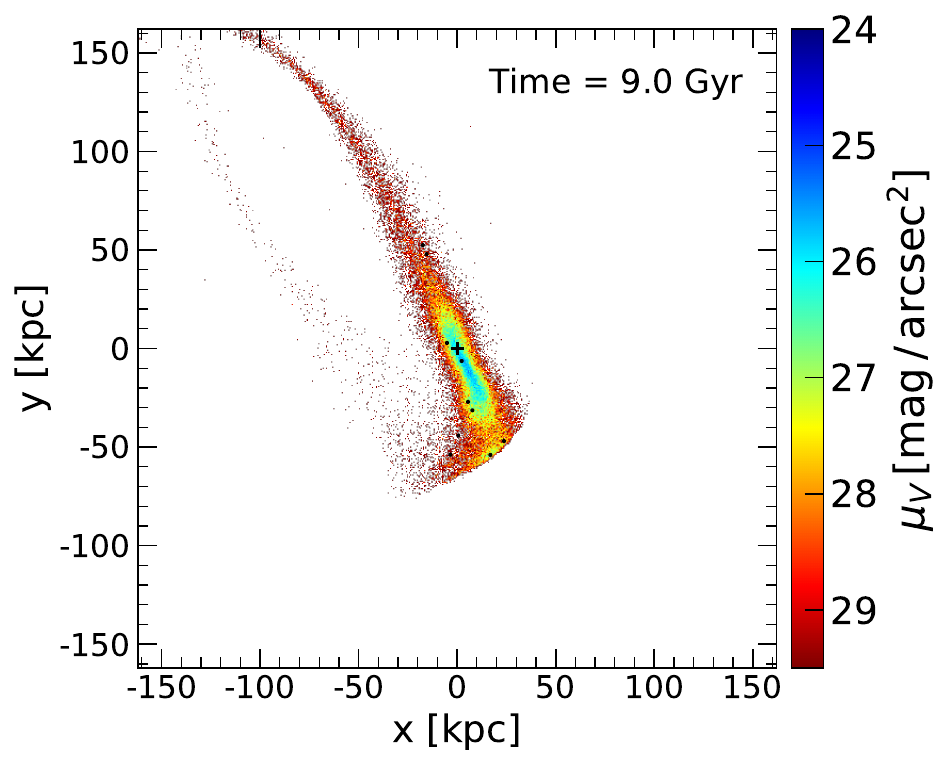}\\
     \includegraphics[keepaspectratio, width=0.9\columnwidth]{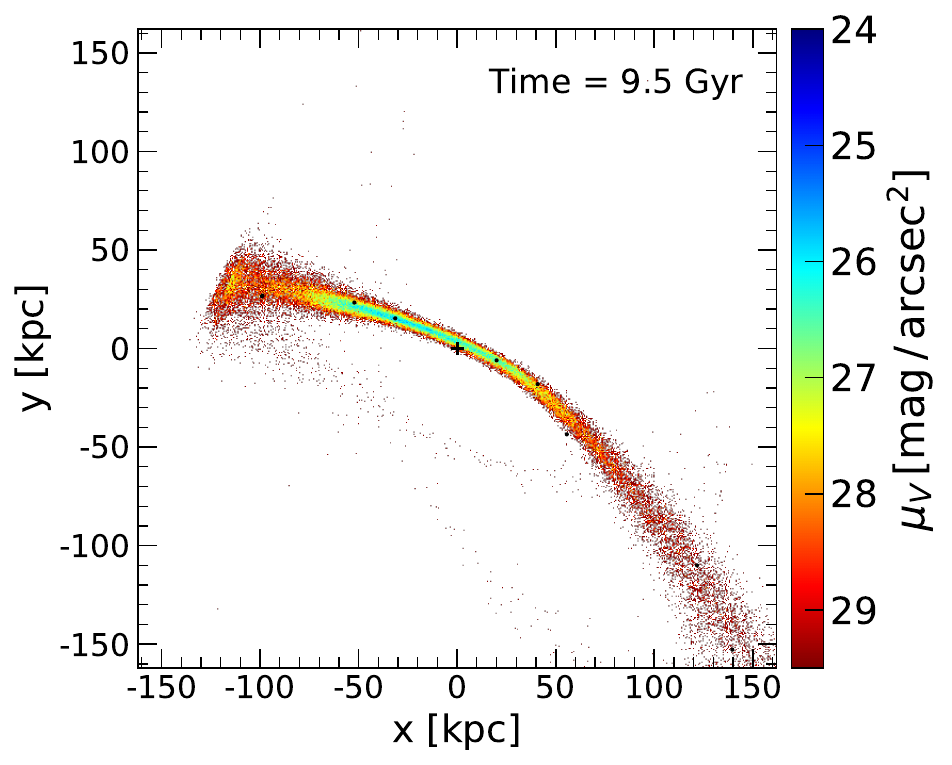}
     \includegraphics[keepaspectratio, width=0.9\columnwidth]{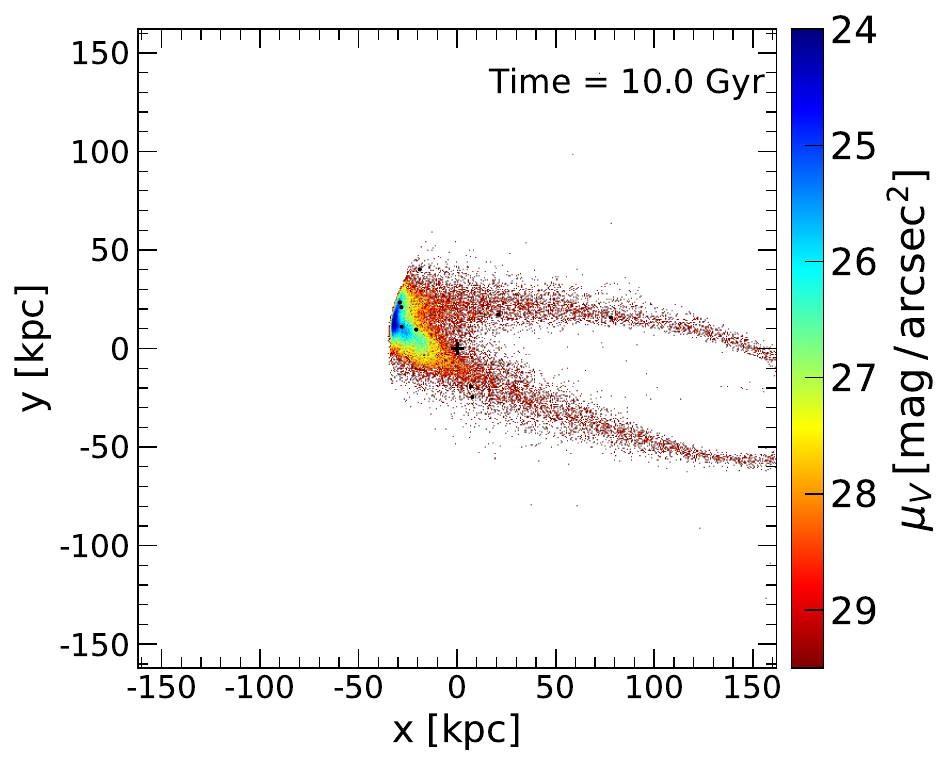}
  \caption{Evolution of surface brightness of the satellite galaxy at $t=7.4$\,Gyr (just before entering DM-deficient phase), 9.0\,Gyr (at the end of DM-deficient phase), 9.5, \&\, 10\,Gyr (after the DM-deficient phase).  In this figure, the origin of the coordinate is shifted to the centre-of-mass of the satellite galaxy.
  }
  \label{fig:SB}
\end{figure*}

For comparison with observations, we show the surface brightness distribution of our simulated satellite galaxy in Fig.~\ref{fig:SB} which was calculated from the projected stellar mass plot assuming a fixed mass-to-light ratio of $M/L_V \sim 3$ \citep{Bruzual03}. 
The surface brightness at the satellite centre is as high as $\mu_V \sim 24$\,mag\,arcsec$^{-2}$, and it smoothly becomes dimmer along the tidal tail down to $29-30$\,mag\,arcsec$^{-2}$. This range of surface brightness is comparable to the observed F8D1 \citep[][see their Fig.\,5]{Zemaitis23}, and we hope to perform further comparison in the future using more realistic cosmological zoom-in simulations with star formation and feedback processes.

% If you want to present additional material which would interrupt the flow of the main paper,
% it can be placed in an Appendix which appears after the list of references.
\color{black}
%%%%%%%%%%%%%%%%%%%%%%%%%%%%%%%%%%%%%%%%%%%%%%%%%%

% Don't change these lines
\bsp	% typesetting comment
\label{lastpage}
\end{document}